\documentclass[aps,prd,twocolumn,amssymb,eqsecnum,nofootinbib,amsmath ]{revtex4}

\usepackage{epsf}  
\usepackage{graphicx}

\begin{document}

\title{Gravitational Radiation from Plunging Orbits \\ 
- Perturbative Study -}

\author{Yasushi Mino}
\email{mino@tapir.caltech.edu}

\author{Jeandrew Brink}
\email{jeandrew@caltech.edu}

\affiliation{
California Institute of Technology, MC 130-33,  Pasadena, CA 91125}

\begin{abstract}
Numerical relativity has recently yielded a plethora of results about kicks from spinning mergers which has, in turn,  vastly increased our knowledge about the spin interactions of black hole systems.  In this work we use black hole perturbation theory to calculate accurately  the gravitational waves emanating from the end of the plunging stage of an extreme mass ratio merger in order to further understand this phenomenon.
This study focuses primarily on  spin induced effects with emphasis on the maximally spinning limit and the identification of possible causes of generic behavior. 
 
We find  that gravitational waves emitted during the plunging phase exhibit  damped oscillatory behavior, corresponding to a coherent excitation of quasi-normal modes by the test particle.
This feature is universal in the sense 
that the frequencies and damping time do not depend on the orbital parameters of the plunging particle. Furthermore, the observed frequencies  are distinct from those associated with the usual free quasi-normal ringing.  
Our calculation suggests that a maximum in radiated energy and momentum  
occurs at spin parameters equal to  $a/M=0.86$ and $a/M=0.81$, 
respectively for  the plunge stage of a polar orbit. The dependence of linear momentum emission on the angle at which a polar orbit impacts the horizon is quantified.  
One of the advantages of the perturbation approach adopted here is that insight into the actual mechanism of radiation emission and its relationship to black hole ringing is obtained by carefully identifying the dominant terms in the expansions used. 
\end{abstract}

\maketitle

\section{Introduction} \label{sec:intro}

As the field of numerical relativity matures, the gravitational wave community gains insight into the most dynamic regions of spacetime \cite{NR}. Possibly the most spectacular strong field event in gravity, namely that of a binary black hole merger, has recently yielded a series of interesting results related to the spins of the holes involved in the collision. A particular highlight is  the discovery of the so-called super-kick configuration, where the spin interaction drastically changes the trajectory of the resultant black hole \cite{skick1,skick2,skick3}.    The body of knowledge relating to spinning compact objects has  greatly increased as a result, inviting analytic explanations. Important analytic strides in this regard have been made \cite{kick, kickPN, kickSP, kickPNe, kickSPe, kickBH}.   
 
The evolution of a black hole binary proceeds through three stages: the inspiral on a quasi stable orbit, the plunge and merger  and the final ringdown stage.
It is in the plunge and merger where the strong field fully nonlinear attraction of general relativity completely dominates the dynamics.

Of these  three stages, the first  and third have been carefully understood analytically. The first stage, during which the slow inspiral over a long timescale results in the  emission of huge amounts of gravitational radiation that dominate the observed signal, has been extensively modeled \cite{PN,BH}.    The final stage, where the highly distorted black hole approaches a Kerr black hole via quasi-normal ringing, is also carefully understood in terms of vacuum black hole perturbation theory  \cite{QNM,QNMfit,QNMWKB,QNMnu}.

The dynamics of the intermediary  phase of plunge and merger, where the binary  transitions from a system described by several parameters related to the two black holes to a resultant system described by far fewer parameters (namely the spin, mass and position of the resultant single black hole), has been less throughly explored. This transition  should by its very nature  display universal behavior. The interest in this phase of the binary evolution is threefold. Firstly, the transition from inspiral to ringdown marks a turning point in the amplitude of the gravitational waveform and is essential for the production of accurate templates for observation. Secondly, it is thought that the plunging phase strongly influences final kick velocity attained by the final black hole. Thirdly, this phase gives us the opportunity to probe highly nonlinear/non-Newtonian events such as horizon formation and the merger process itself, to which currently only numerical relativity gives access. 

The first issue can be addressed by attempting to extrapolate waveforms produced for the first and third stages to bridge the chasm of understanding present in the second stage. A method which extends the post-Newtonian  (PN) approximation into this regime while leaning on numerical relativity results was proposed in \cite{kickPNe,kickSPe}, however it is mentioned there that the uncertainty due to the plunging stage remains considerable.
The amount of linear momentum emitted during the plunge stage is expected to be significant in comparison to that emitted during inspiral,  making the contribution from the final plunge stage to the final velocity of the kicks of the black hole important. Furthermore,  this statement also implies that the position in the orbit at which a binary merges strongly influences the end result.
 While the  kick velocities resulting from spin interactions can be  modeled effectively by fitting formulas which treat the strong field interactions as an effective black box \cite{skick1, skick2, skick3, kickSP}, a careful understanding of the interactions that produce them and the precise connection to gravitational radiation still remains somewhat illusive.  A constructive formalism for treating and understanding strongly nonlinear effects in the highly dynamic regions of spacetime during the plunge and  merger phases is in its early stages of development. Subtle details, exhibited in the waveform and dynamics during the merger phase, such as the anti-kick observed by \cite{akick} give us further clues as to  dynamics that require explanation.    As numerical relativity now takes the lead in ushering in a new stage in our analytic exploration of the Einstein field equations, we can begin to build our intuition about how  interactions between compact object  take place and develop the tools for describing them.

In this paper, we study gravitational waves emitted from the end of the plunging phase of a small black hole  falling into a spinning black hole. The purpose of the calculation is to obtain a physical understanding of the origin of the radiation emitted and to identify possible causes of universal behavior. 
We explore the extreme mass ratio case via black hole perturbation theory. This approach is used for two reasons. Firstly, it makes the problem analytically tractable at minimal computational cost. Secondly, it gives insight into the physical mechanism by which the radiation is produced by the plunging phase of the black hole merger and directly relates that picture  to the resulting gauge invariant radiation quantities computed at infinity.

The proposed model allows the  effect of spin to be explored by considering the  background spacetime to be a Kerr black hole with mass $M$ and spin $a$. The second smaller black hole is modeled as a point particle moving on a geodesic orbit described in Sec.\ref{sec:geo}.  Past studies in which  black hole perturbation theory is applied to a Schwarzschild black hole have indicated that an important contribution to the energy radiated during an infall has its origin just outside the event horizon \cite{plunge1}. It is on this region we will focus. Since we are interested in the plunging phase of the orbit, we make  use of the near horizon approximation, which is introduced in Sec.\ref{sec:geo}.  Features of the final stage of the plunge that influence the emitted radiation are identified in Sec.\ref{sec:mp} and App.\ref{app:Source}. The Teukolsky formalism is used to compute the gravitational waveform from the end of the plunging stage in Sec.\ref{sec:mp}, and to compute the energy and momentum fluxes in Secs.\ref{sec:ene} and \ref{sec:mom} respectively.
 Numerical explorations on a Kerr background  within the perturbation framework have indicated that spin enhances the radiation emission \cite{plunge3,plunge5}.  This was understood in the   context of the lower damping rates of the quasi-normal modes in the presence of spin\cite{QNM}.
In our work, the spin enhancement effect is  carefully quantified analytically during the end of  the  plunge phase.  

Our work aims at analytically quantifying features of the transition from the plunge to the  ring-down  phase which cannot be captured by extending purely post-Newtonian results \cite{plunge2}.   
The analysis presented here does not extend into the regime of free quasi-normal ringing  observed when a highly distorted black hole approaches Kerr in the absence of an external source. This is due to the fact that the  perturbing  second black hole is still present  and therefore a source of waves.

 A possible concern about  modeling the end of the black hole plunge  in the extreme mass ratio limit with first order  perturbation theory is that the problem has been oversimplified by ignoring non-linear perturbations and employing the point particle approximation for the smaller black hole. However,  it should be observed that the dynamical timescale associated with the small black hole is comparatively short, making the equilibration process of internal perturbations of the small black hole rapid and thus justifying the point particle approximation. A huge advantage of the model is that it takes into account  the fully relativistic frame dragging effect on the orbit in the vicinity of the event horizon. 
A further advantage is that it allows for the exploration of  the maximally spinning black hole limit. This is a case  of particular theoretical interest which  can only be explored by taking a limit within this analytic perturbation framework.  Finally,
 the perturbation formalism implicitly takes into account the horizon deformation of the background spacetime, although care has to be exercised when dealing with a source that crosses the event horizon, as is shown in App.\ref{app:teu0}. It is hoped that with the rapid advance of numerical simulations into the extreme mass ratio regime,  the predictions of our model will soon be numerically tested.

 The analysis presented in this paper allows us to identify the following effects. At the end of the plunging stage just before the small black hole passes through the event horizon, the radiation is dominated by a peculiar feature indicative of the ergoregion. This feature is best described as follows. In its own reference frame, the small black hole passes through the event horizon in finite time; however, to a distant observer it remains a radiating mirage  stuck on the event horizon and dragged into motion at a tremendous speed, emitting gravitational waves.  In practice, it is not the mirage emitting gravitational waves, but the nonlinear perturbed gravitational field that has stored up energy and angular momentum and is slowly leaking them out to the observer at infinity. The frequency and damping rate display universal behavior in that the earlier details of the orbital trajectory are irrelevant;  only the position at which the small black hole crosses the horizon matters. The emitted radiation has a frequency and damping rate analogous to quasi-normal ringing but at a frequency distinct from it.
Pretorius \cite{NR} gave a heuristic description of the sinusoidal bobbing motion observed in the numerical simulations of two black holes in the so-called super kick configurations in terms of a frame dragging phenomenon.  He mentions that the position in the orbit where the merger takes place ultimately determines the magnitude and direction of the kick. While the magnitude of the kick cannot be accurately accounted for by the extreme mass ratio  analysis 
of the plunge phase of the orbit, 
the dependence of the radiated linear momentum on the angular position on the horizon at which the small black hole enters can, and is calculated in Sec.\ref{sec:mom}. We conclude the paper with possible observational and analytic  implications in Sec.\ref{sec:con}.  

Throughout this paper,   geometrized units, 
such as \mbox{$G=c=1$} are used.

\section{Orbital description of plunging object 
} \label{sec:geo}
In this section we describe the nature of the orbit in the final plunging phase.  The rapid timescale on which the dynamics of the plunging phase occurs, implies that the orbital evolution of the small black hole is well approximated by a geodesic in the background Kerr geometry, ignoring the radiation reaction effect. Since our focus is on the late plunging phase,  we then further specialize the orbit to the near horizon region, also referred to as the near-horizon limit.

The geodesic equations describing the trajectory of a freely falling test particle in the Kerr geometry can be expressed in   Boyer-Lindquist coordinates 
$\{t,r,\theta,\phi\}$ using the equations

\begin{align}
\left({dr \over ds}\right)^2 
&= 
 R(r) 
 \,, & 
\left({d\theta \over ds}\right)^2 
&=
 \Theta(\theta) 
\label{eq:geo_r} \,, 
\end{align}
where \begin{align}
 R(r) &= \left[E(r^2+a^2)-aL_z\right]^2-\Delta\left[(a E-L_z)^2+r^2+C\right] \notag\\
 \Theta(\theta)&=C-\cos^2\theta\left\{a^2(1-E^2)+\left(L_z^2 \over \sin^2\theta\right)\right\} \notag
\end{align}
and
\begin{align} 
{d\phi \over ds} 
&= -\left(aE-{L_z \over \sin^2\theta}\right)
+{a \over \Delta}\left(E(r^2+a^2)-aL_z\right)
\label{eq:geo_phi} \,, \\ 
{dt \over ds} 
&= -\left(aE-{L_z \over \sin^2\theta}\right)a\sin^2\theta\notag \\
& \ \ \ \ \ \ \ \ \   +{r^2+a^2 \over \Delta}\left(E(r^2+a^2)-aL_z\right)
\label{eq:geo_t} \,. 
\end{align}
In the above expressions, the constants of motion for a particular orbit are denoted $E$, $L_z$ and $C$ for energy, azimuthal angular momentum and Carter constant respectively; furthermore, the function $\Delta$ is defined to be $\Delta = r^2-2Mr+a^2$ 
and the affine parameter $s$ 
is related to the proper time $\tau$ 
by $s = \int d\tau/\Sigma$, with $\Sigma=r^2+a^2\cos^2\theta$.

To facilitate understanding of the radiation content emanating from the orbit just before the test mass falls into the horizon, we further specialize these equations to the near horizon limit.
Denote the radial position of the event horizon by $r = r_+  = M+\sqrt{M^2-a^2}$. The quantitative features of an in-falling orbit in the near horizon region  ($r \approx r_+$), can be adequately described by approximating the geodesic equations  (\ref{eq:geo_r})-(\ref{eq:geo_phi}) by
\begin{align}
{dr \over ds} &= -2Mr_+(E-\Omega_H L_z) +O(r-r_+) 
\label{eq:geo_r1} \,, \\ 
{d\theta \over ds} &= \pm\sqrt{\Theta_0} +O(\theta-\theta_0) 
\,,\quad 
\Theta_0 = \Theta(\theta_0) 
\label{eq:geo_the1} \,, \\ 
{d\phi \over ds} 
&= {a\over \kappa}(E-\Omega_H L_z){r_+ \over r-r_+}
+O((r-r_+)^0)
\label{eq:geo_phi1} \,, \\ 
{dt \over ds} 
&= {2Mr_+\over\kappa}(E-\Omega_H L_z){r_+ \over r-r_+}
+O((r-r_+)^0)
\label{eq:geo_t1} \,, 
\end{align}
where
$\Omega_H={a \over 2Mr_+}$ is the horizon's angular velocity, 
and $\theta_0$ is the polar angle 
at which  the particle falls into the horizon, 
and the dimensionless constant $\kappa = \sqrt{1-(a/M)^2}$. 

The geodesic 
in the near-horizon limit has the following analytic solution; 
\begin{align}
r &= r_+ \left(1+e^{-\kappa(t-t_0)/r_+}\right)
\,,
&{dr \over dt}  &= -\kappa {r-r_+ \over r_+} 
\,, \label{eq:geo_r2} \\ 
\theta &= \theta_0 
\mp{\sqrt{\Theta_0} e^{-\kappa(t-t_0)/r_+}   \over 2M(E-\Omega_H L_z)}
\,, 
&{d\theta \over dt} 
&= \pm{\kappa\sqrt{\Theta_0}(r-r_+  ) \over 2Mr_+^2(E-\Omega_H L_z)}
\,, \label{eq:geo_the2} \\ 
\phi &= \Omega_H(t-t_0) +\phi_0 
\,, 
&{d\phi\over dt} &= \Omega_H
\,, \label{eq:geo_phi2} 
\end{align}
where $t_0$ and $\phi_0$ are  integration  constants. 
This approximation holds for $t-t_0 >> r_+$. 

Note that the well-known gravitational time dilation effect is clearly manifest in terms of the singular behavior of Eq.(\ref{eq:geo_t1}). While the particle passes through the event horizon in finite proper time in its own local reference frame, it appears to a distant observer who describes the particle motion in terms of coordinate time to be slowing down infinitely as it approaches the event horizon. As a result, the motion in the  radial and polar directions becomes frozen onto the event horizon at their entering positions. If the central black hole has no spin, this particle will cease to be a source of gravitational waves during the final stages of the plunge to the order of approximations made.

If the central black hole is spinning, however, Eq.(\ref{eq:geo_phi2}) indicates that the motion in the azimuthal angle does not slow down. Instead the frame dragging effect ``pulls'' the particle into motion around the black hole with an azimuthal angular velocity that approaches the horizon angular velocity.  The particle motion around a spinning black hole in the final stage of the plunge is thus, from the perspective of a distant observer, entirely dominated by this azimuthal motion which is insensitive to the previous details of the orbital evolution and characterized only  by the mass and spin of the central object.  This effect is highly relativistic, it cannot be captured by an analysis based on a post-Newtonian expansion, and is the cause of strong gravitational wave emission during the plunging phase of a spinning black hole.  This statement will be made explicit in Sec.\ref{sec:mp}.

\begin{figure*}
\includegraphics[width=2 \columnwidth]{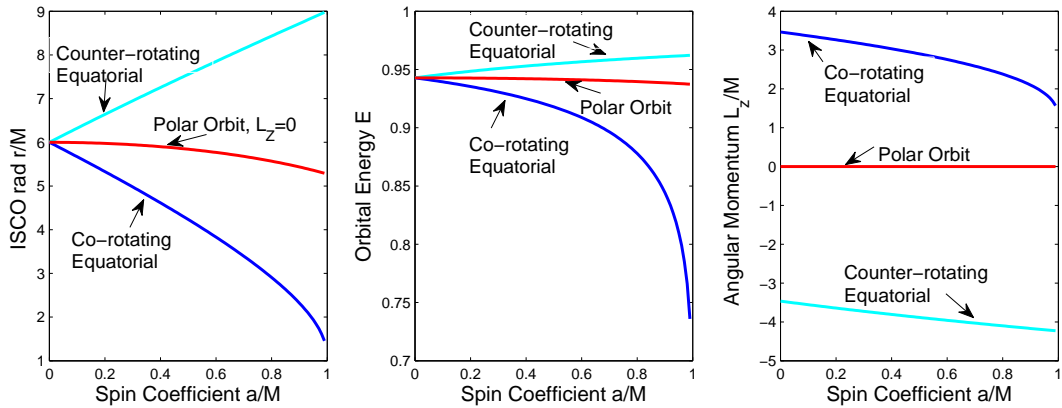}
\caption{ISCO radius, orbital energy ($E$) and azimuthal angular momentum ($L_z$) as a function of the spin coefficient $q=a/M$}
\label{pic:ee}
\end{figure*}

While the constants such as $E$, $L_z$ and $C$ describing orbital motion do not enter directly into the description of the final stages of the plunging orbit itself,  they do make their appearance in the dominant terms of the perturbation theory (see App.\ref{app:teu0}) and are thus required for explicit evaluation. To this end, we have selected three plunging orbits for explicit evaluation: 
the co- and counter-rotating plunging trajectories from the innermost stable circular orbit (ISCO) in the equatorial plane (i.e. $C=0$ and $\theta_0=\pi/2$), and the orbit plunging from the polar ISCO  with vanishing $z$-component of the angular momentum, (i.e. $L_z=0$).  The constants of motion characterizing these orbits can be calculated by writing down the condition that the orbit remain circular, namely by requiring  that $R(r,E,L_z,C ) = dR(r,E,L_z,C) / dr  =0$ at the ISCO radius $r=r_{ISCO}$. The additional condition that the orbit remain stable adds the equation $d^2R(r,E,L_z,C)  / dr^2 = 0$. If one further assumes that the orbit is either polar or equatorial, namely that $C=0$ or $L_z=0$, then the three equations allow the three unknowns, $E$, $L_z$ and $r_{ISCO}$, to be expressed in terms of the black hole mass $M$ and spin $a$.

Fig.\ref{pic:ee} depicts the ISCO radius $r_{ISCO}$, orbital energy $E$ and azimuthal angular momentum $L_z$ as functions of the spin coefficient $q=a/M$.  Due to spin-orbit coupling, the orbital energy of the ISCO depends strongly on the spin coefficient when the orbit lies in the equatorial plane.When the orbit is co-rotating/counter-rotating, the spin-orbit coupling works as a repulsive/attractive force that stabilizes/destabilizes the orbit with respect to  the gravitational attraction of the background black hole. 
As a result, the radius of the co-rotating/counter-rotating ISCO becomes smaller/larger when compared to the ISCO in the non-spinning limit. This effect is displayed in the first panel of  Fig.\ref{pic:ee}. 
For the polar orbit which corresponds to the super kick configuration, the orbital energy of ISCO depends weakly on the spin coefficient. It turns out that only the $E$ and $L_z$ orbital parameters ultimately enter the waveform calculation  in Sec.\ref{sec:mp}. However for completeness,  the Carter constant $C$ in the case of the plunge from the  polar ISCO orbit  can be found in terms of energy $E$ and $r_{ISCO}$ to be
 
\begin{eqnarray}
C&=& E^2 \frac{(a^2+r^2_{ISCO})^3(3r^2_{ISCO}-a^2)}{8(r^2_{ISCO}-a^2)^3}.\notag
\end{eqnarray}

\section{Gravitational waves emitted during the final stage of the plunge} \label{sec:mp}
The gravitational waves emitted at infinity can be computed using the black hole perturbation theory expressed using the Teukolsky formalism.  The radiation contained in the resulting waves is expressed in terms of the Newman-Penrose curvature scalar  $\Psi_4$:   
\begin{eqnarray}
\Psi_4(r \to \infty)
&=& {1 \over 2}\left(\ddot h_+-i\ddot h_\times\right)
\,, 
\end{eqnarray}
where $h_{+/\times}$ represents the wave polarization 
and the double dot indicates two time-derivatives. 

Initially it may appear that, due to its bewildering complexity, the  full waveform calculation of a plunging orbit within the context of black hole perturbation theory may  best be attempted numerically by means of direct integration of the Teukolsky equation as was done in \cite{plunge1,plunge3,plunge4,plunge5}. However, it turns out that identifying the dominant terms  and spin effects 
  in the calculation analytically is particularly insightful. Furthermore,  in the rapidly spinning case, it is highly likely that numerics will not be able to capture the final stages of the plunge and that this regime is thus only accessible via analytic perturbation theory.
The full Teukolsky machinery is summarized in gory detail in App.\ref{app:teu0}. This  section highlights the aspects that make the problem tractable analytically, the approximations made, and the results obtained. 

  The scalar $\Psi_4$ can be expanded in terms of its  Fourier-Harmonic components as 
\begin{align}
\Psi_4 &=& {1 \over (r-ia\cos\theta)^4} 
\int d\omega \sum_{\ell m} e^{-i\omega t +im\phi} 
R_{\ell m\omega}(r) S_{\ell m\omega}(\theta) 
\,. \label{eq:cuv11} 
\end{align}
where $R_{\ell m\omega}(r)$  and $S_{\ell m\omega}(\theta)$  are the radial and angular Teukolsky functions respectively.   The influence of the motion of the point particle as it plunges into the black hole enters the radial Teukolsky equation as a source term.  The particular solution to the radial Teukolsky equation can be found by means of evaluating an integral containing a Green's function constructed from an expansion of homogeneous solutions to the Teukolsky equation. The fact that we are only interested in the solution at $r \rightarrow \infty$  simplifies the expression  given in  App.\ref{app:teu0}  to 
\begin{align}
R_{\ell m\omega}(r\to\infty) &\to
{r^3 e^{i\omega r^*} \over 2i\omega B^{in}_{\ell m\omega}} 
\int^\infty_{r_+} dr 
R^{in}_{\ell m\omega} \Delta^{-2}T_{\ell m\omega} \notag\\
&\to r^3 e^{i\omega r^*} Z_{\ell m\omega} 
\,, \label{eq:teuk_r11} 
\end{align}
where $ B^{in}_{\ell m\omega}$ is a mode dependent constant,  $T_{\ell m\omega} $ the source term related to the stress energy tensor describing the particle motion and $R^{in}_{\ell m\omega}$  is the homogeneous in-going Teukolsky function with asymptotic behavior
\begin{align}
R^{in}_{\ell m\omega} &\to& \left\{ \begin{matrix}
B^{trans}_{\ell m\omega}\Delta^2 e^{-ikr^*} 
& \hbox{for $r\to r_+$} \\
B^{ref}_{\ell m\omega} r^3e^{i\omega r^*} 
+B^{in}_{\ell m\omega} r^{-1}e^{-i\omega r^*} 
& \hbox{for $r\to\infty$} 
\end{matrix}
\right. 
\, . \label{eq:in_r_teu11} 
\end{align}
In the above equation, $k=\omega-ma/2Mr_+$ 
and $r^*$ is the tortoise coordinate defined by 
$(dr^*/dr)=(r^2+a^2)/\Delta$.

 The nature of the problem,  with the point particle on the final stages of the plunging orbit, requires that this source integral need only be evaluated  near the horizon to capture the essentials features of the the radiation emitted during this phase. Thus the asymptotic form of  $R^{in}_{\ell m\omega}$ given in Eq.\eqref{eq:in_r_teu11} can be used. It is useful to observe that in this region $\Delta \approx 2 \kappa M (r-r_+) $.

The dominant contribution to the source term results 
 the rapid frame dragging motion observed when the particle is frozen onto the horizon at the angle  of impact $\theta_0$. This contribution is contained in
 the $T_{\overline{m}\overline{m}}$ component of the stress energy tensor projected along the tetrad legs, defined in App.\ref{app:teu0}.  The dominant scalings of the other components of the source term on the event horizon for a plunging orbit are written out in full in App.\ref{app:Source}.

With these approximations, the function $Z_{\ell m\omega}$ entering the asymptotic form 
of the the radial Teukolsky function (\ref{eq:teuk_r11}) can be evaluated to be  
\begin{align}
Z_{\ell m\omega} &= \tilde Z_{\ell m\omega} 
\int dt e^{i\omega t-im\phi}
{r-r_+ \over r_+} e^{-ikr^*}
\left\{1+O\left(\epsilon_H)\right)\right\}
\,, \label{eq:zz0} \\ 
\tilde Z_{\ell m\omega} &=
\mu\sqrt{2 \over\pi}{1 \over 2i\omega}
{B^{trans}_{\ell m\omega} \over B^{in}_{\ell m\omega}} 
{\kappa^3 M \over r_+ (E-\Omega_H L_z)} \notag\\
& \quad \quad
\times \left[1-2i{kr_+ \over \kappa}
-2\left({kr_+ \over \kappa}\right)^2\right]
\nonumber \\ & \times 
\left(aE-{L_z \over \sin^2\theta_0}\right)^2
{r_+-ia\cos\theta_0 \over r_++ia\cos\theta_0}
\sin^2\theta_0 S_{\ell m\omega}(\theta_0)
\,. \label{eq:zz1} 
\end{align}
where $\epsilon_H= (r-r_+)/ r_+$.

Using the solution to the geodesic equations specialized to the region near the event horizon (\ref{eq:geo_r2})-(\ref{eq:geo_phi2}) to evaluate $\phi$ along the orbit
and $r^* \to (r_+/\kappa)\ln(r-r_+)$, 
the $t$-integral of (\ref{eq:zz0}) becomes 
\begin{align}
\int^\infty_T dt &\  e^{i\omega t-im \Omega_H t}
e^{-(\kappa/r_+)(t-t_0)}e^{ik(t-t_0)}\notag\\
&= {i \over 2}e^{ik(2T-t_0)-(\kappa/r_+)(T-t_0)}
{1 \over \omega -m\Omega_H+i{\kappa \over 2 r_+}} 
\,, \label{eq:zz2} 
\end{align}
where the lower bound of the integration, $T$, 
must be chosen so that  $e^{-(\kappa/r_+)(T-t_0)}<<1$, 
thus, ensuring the validity of the near-horizon expansion.

From Eq.\eqref{eq:zz2} it becomes clear that the emitted radiation is strongly peaked at the frequency $\omega =m\Omega_H-i{\kappa \over 2 r_+}$. Taking this into account, the asymptotic form of the curvature scalar $\Psi_4$
is 
\begin{align}
\Psi_4(r \to \infty) = -\pi
{e^{\kappa t_0 \over 2r_+} \over r} 
\sum_{\ell m} \left[
e^{-i\omega (t-r^*) +im\phi}S_{\ell m\omega}(\theta)
\tilde Z_{\ell m\omega} 
\right]
\,, \label{Psi4RAD} 
\end{align}
where the sum is evaluated at an angular frequency of $\omega =m\Omega_H-i{\kappa \over 2 r_+}$.
This result represents a decaying spectrum of gravitational waves with frequency around $\omega =m\Omega_H$ being damped  in a manner that admits a quality factor of 
\begin{eqnarray}
Q = {|m| \over \sqrt{1-(a/M)^2}}\left({a \over M}\right) 
\,. \label{eq:qfac} 
\end{eqnarray}

The analytic results characterizing the gravitational waves  from the plunging phase of the orbit, yields the following observations. 
Firstly,  the frequency and quality factors of the observed radiation differ from the well-known 
quasi-normal ringing modes of black hole perturbation theory. The quasi-normal ringing that is observed and expected to follow most binary black hole collisions can be  characterized  as damping modes associated with vacuum perturbations.  The quasi-normal ringing is often associated with the relaxation of perturbations of the stationary black hole horizon.
 The origin of the radiation described by Eq.\eqref{Psi4RAD}  is somewhat different. It is produced by  perturbations coherently induced by the particle still present on the horizon.  The damping behavior is primarily due to the fact that the particle falls through the horizon and ceases to be a source of perturbation.
For a quantitative comparison with the least damped ($l=m=2$) quasi-normal mode, one can make used of a fitting formula suggested by  Echeverria \cite{QNMfit} that gives the quality factor and frequency of this mode as a function of spin to be
\begin{align}
Q_{QN} &= 4(1-a)^{-0.45} f_Q(a), \\
\omega_{QN} &= [1-0.63(1-a^{0.3})]f_f(a), 
\end{align}
where $f_Q$ and $f_a$ are functions of  order unity as the spin factor ranges from 0 to 1, i.e. $f_Q \in [1.05,0.95]$ and  $f_a \in [1.02,0.97]$.
Note that for spin factor $a/M<0.97$,  both the frequency and quality factor of the quasi-normal mode exceed those predicted by the driven motion of the plunging phase.  As a result, the moment the particle ceases to be observable and thus to serve as a source driving the radiation, the quasi-normal ringing effect will rapidly begin to dominate the observables.
In the region of very high spin  ($a/M>0.97$) the two effects of quasi-normal ringing and radiation driven by particle motion on the horizon become almost indistinguishable.  
In the low spin limit the difference between the frequency and quality factor of the quasi-normal ringing and those associated with the radiation during plunge phase begin to differ considerably. However, it should also be noted that in this limit, our  assumption that the dominant term in the stress energy tensor expansion is the frame dragging induced rotation breaks down.

Secondly,
 while the small particle is plunging, the radiation emitted broadcasts features of the background spacetime, namely Kerr, rather than features reminiscent of its orbital trajectory. The spectrum peak at  around $\omega \sim m\Omega_H$ is determined entirely by the constants describing the Kerr black hole without any reference to the orbital constants of the plunge.
As argued in Sec.\ref{sec:geo}, 
the frame-dragging effect accelerates the rotational velocity 
of the particle around the spin axis of the black hole. 
The origin of the universal value of the frequency peak can thus best be explained as the frame dragging  effect  compensating for the  gravitational time dilation effect of the radiation. 

Thirdly, the quality factor associated with the radiation emitted during the plunging phase approaches infinity in the maximally spinning limit. The radiation emitted during the plunge in this limit may  provide a sensitive means of estimating the spin of a rapidly rotating black hole. The quality factors for the radiation of gravitational waves with ($m=2$) for a series of spin parameters are tabulated below.

\begin{center}
\begin{tabular}{|c||c|c|c|c|c|c|c|}
\hline
$a/M$ & 0.95 & 0.97 & 0.98 & 
0.99 & 0.995 & 0.998 & 0.9999 \cr 
\hline
$Q$ & 6.08 & 7.98 & 9.85 & 
14.04 & 19.92 & 31.58 & 141.41 \cr  
\hline
\end{tabular}
\end{center}
The large quality factor obtained for the rapidly spinning limit can be used to explain the enhanced  
emission of gravitational energy and momentum 
from the plunge observed in the numerical studies made within the Teukolsky perturbation framework   
\cite{plunge3,plunge5}.
The feature that the quality factor approaches  infinity in the maximally spinning limit may appear to endanger the assumptions made earlier regarding the insignificance of radiation reaction on the orbit. For if a  mode at a finite frequency lasts an infinitely long time, it will radiate an infinite amount of energy. It should be noted that the quasi-normal modes of Kerr display a similar feature. In the past,  it was feared that this infinite quality factor in the case of quasi-normal modes implied the instability of the Kerr black hole, however this turns out not to be the case \cite{QNM,QNMWKB}.  The resolution of this conundrum  with respect to the current problem is explored in the Sec.\ref{sec:ene}

Fourthly, the relevant expansion parameter near the horizon of the black hole is $\epsilon_H=(r-r_+)/r_+$. In the analysis performed in this section, only the leading order terms were retained. This approach is now justified by the following argument. Suppose one goes one step further, and computes the higher order  terms of the orbit and stress energy. This will give rise to additional terms 
proportional to 
$\int^\infty dt \ e^{i\omega t-im \Omega_H t}
e^{-n(\kappa/r_+)(t-t_0)}e^{ik(t-t_0)}
\propto {1 \over \omega -m\Omega_H+i n(\kappa / 2 r_+)}$ 
which enter the calculation for the radial Teukolsky function, Eq.(\ref{eq:zz0}), where $n=2,3,\cdots$. 
This suggests that the curvature perturbation 
in the plunging phase can more generally be written as 
\begin{align}
\Psi_4(r \to \infty) = 
{1 \over r}\sum_{\ell m, n=1,2,\cdots} \left[
e^{-i\omega (t-r^*) +im\phi}S_{\ell m\omega}(\theta)
\tilde X_{\ell m \omega} 
\right]
\,. \end{align} 
where the  frequency $\omega$ entering the expression is different for every mode and  given by ${\omega =m\Omega_H-in{\kappa \over 2 r_+}}$. 
Note that in the above equation, the damping factor has been enhanced by a factor of $n$ and so the dynamics described thus far are truly the dominant behavior,  since all higher order terms are more strongly damped.

\section{Energy Flux 
} 
\label{sec:ene}

The picture presented thus far is that of a small body falling into the central black hole on a dynamical timescale with respect to proper time, which  for sufficiently extreme mass ratios, is assumed to be  well short of the radiation reaction timescale. This justifies the  assumption that gravitational radiation backaction on the orbit is negligible and that the particle plunges on a geodesic trajectory.

In Eq.(\ref{eq:qfac}) of Sec.\ref{sec:geo},  it was shown  that for rapidly spinning black holes, the quality factor becomes very large in the maximally spinning limit. This fact could possibly indicate that the gravitational wave emission at the horizon angular frequency for a maximally rotating black hole  may be considerable and last for a long period of time. If this were the case, it would nullify the assumptions previously made regarding the geodesic trajectory of the plunging black hole. 

In this section, we calculate the gravitational energy flux during the final stage of the plunge and estimate the maximum amount of energy the particle can emit during the plunging phase. In doing so, we check the validity of the assumptions made and explore the possibility of direct detection of this stage of the orbit.

The gravitational energy flux at infinity can be expressed as
\begin{eqnarray}
{dE \over dt} 
&=& \lim_{r \to \infty} 
\left[{r^2 \over 16\pi}\int d\Omega 
\left| \int^t dt \Psi_4 \right|^2 \right] 
\nonumber \\ 
&\to& {\pi^2 \over 8} e^{-{\kappa \over r_+}(t-t_0-r^*)}
\sum_{\ell m} {\left|\tilde Z_{\ell m \bar\omega_m}\right|^2 
\over |\bar\omega_m|^2} 
\,,
\end{eqnarray}
where  $\bar\omega_m=m\Omega_H-i{\kappa \over 2r_+}$ is used. 
The total energy radiated from $t-r^*=T$ to $t-r^*=\infty$ 
is given by 
\begin{align}
\Delta E
&= {\pi^2 \over 8} e^{-{\kappa \over r_+}(T-t_0)}
{r_+ \over \kappa} 
\sum_{\ell m}{\left|\tilde Z_{\ell m \bar\omega_m}\right|^2 
\over |\bar\omega_m|^2} 
\nonumber \\ 
&= {\pi \over 16} \mu^2 
{\kappa^5 M^2 \over r_+} e^{-{\kappa \over r_+}(T-t_0)}
{\left(aE-{L_z\over\sin^2\theta_0}\right)^4 
\over (E-\Omega_H L_z)^2} \sin^4\theta_0 \notag\\
&  \quad \quad\times \sum_{\ell m}{\left|S_{\ell m \bar\omega_m}(\theta_0)\right|^2 
\over |\bar\omega_m|^4} 
\left|{B^{trans}_{\ell m \bar\omega_m}
\over B^{in}_{\ell m \bar\omega_m}}\right|^2 
\,. \label{eq:rad_E0} 
\end{align}
Eq.(\ref{eq:rad_E0}) can be evaluated by using the  analytic expressions 
for the homogeneous Teukolsky functions in the low frequency limit given in  App.\ref{app:teu1}. It can be observed that the relative contribution of higher $\ell$-modes converge very rapidly; in fact, they fall off super-exponentially, allowing us to consider only the   $\ell=2$ modes. So doing we have
\begin{align}
\Delta E
&\approx {\pi \over 1024} \mu^2 
{\kappa^{11} M^2 \over r_+^7} e^{-{\kappa \over r_+}(T-t_0)}
{\left(aE-{L_z\over\sin^2\theta_0}\right)^4 
\over (E-\Omega_H L_z)^2} \sin^4\theta_0 
\nonumber \\ & \times 
\Biggl\{{8 \over 45}
+\left({32 \over 15}-{8 \over 5}\sin^2(\theta_0)\right)
\left({a \over \kappa M}\right)^2
 \notag\\
&\quad \quad +\left({128 \over 15}-8\sin^2(\theta_0)
+{4 \over 5} \sin^4(\theta_0)\right)
\left({a \over \kappa M}\right)^4
\nonumber \\ & \qquad
+\left({512 \over 45}-{56 \over 5}\sin^2(\theta_0)
+{4 \over 3} \sin^4(\theta_0)\right)
\left({a \over \kappa M}\right)^6 \Biggr\}
\,. \label{eq:rad_E} 
\end{align}
In the maximally spinning limit, the scaling behavior of the total radiated energy with respect to the quality factor can be expressed as
\begin{eqnarray}
\Delta E \propto \kappa^5 \propto Q^{-5} 
\,. 
\end{eqnarray}
This suggests that
even though gravitational waves could be emitted 
for an extremely long time at a finite frequency 
in the maximally spinning limit, 
the wave amplitude of this radiation is suppressed  
and the resultant total energy radiated small. 
This shows conclusively that if the mass ratio is sufficiently extreme,  the backaction 
due to the emission of gravitational waves  is negligible. 
The approximation of the spacetime geometry 
by the linear metric perturbation theory remains valid, as does the assumption that the in-falling particle moves on a geodesic orbit. 
This result further implies that it will be difficult to observe
highly relativistic dynamics of the particle orbiting just outside the horizon in the maximally spinning case
because of the strong suppression of the wave amplitude.

The observed suppression of the wave amplitude in the maximally spinning case can be explained using the radial Teukolsky equation (\ref{eq:teu}). Near the horizon the homogeneous radial  Teukolsky equation (\ref{eq:teu}) becomes
\begin{align}
(r-r_+){d \over dr}\left({1\over r-r_+}
{dR_{\ell m\omega}\over dr}\right)
+{(kr_+)^2+2i\kappa(kr_+) \over \kappa^2(r-r_+)^2}
R_{\ell m\omega} = 0 
\,. 
\end{align}
The pole in the potential has a coefficient of  $\kappa^{-2}$ . As  a result, in the maximally rotating limit  ($\kappa \to 0$) it becomes increasingly difficult for radiation originating from a source near the horizon to propagate outward and reach infinity. The potential barrier in the Teukolsky equation thus produces a suppression effect that competes with the high quality factor of emitted radiation described in Sec.\ref{sec:geo} and ultimately dominates as the black hole becomes maximally spinning. The competition between these two effects result in a spin coefficient less than unity at which the maximum amount of energy is radiated during a plunge. This feature is present in all orbits explored. 

\begin{figure*}[t]
\includegraphics[width=2 \columnwidth]{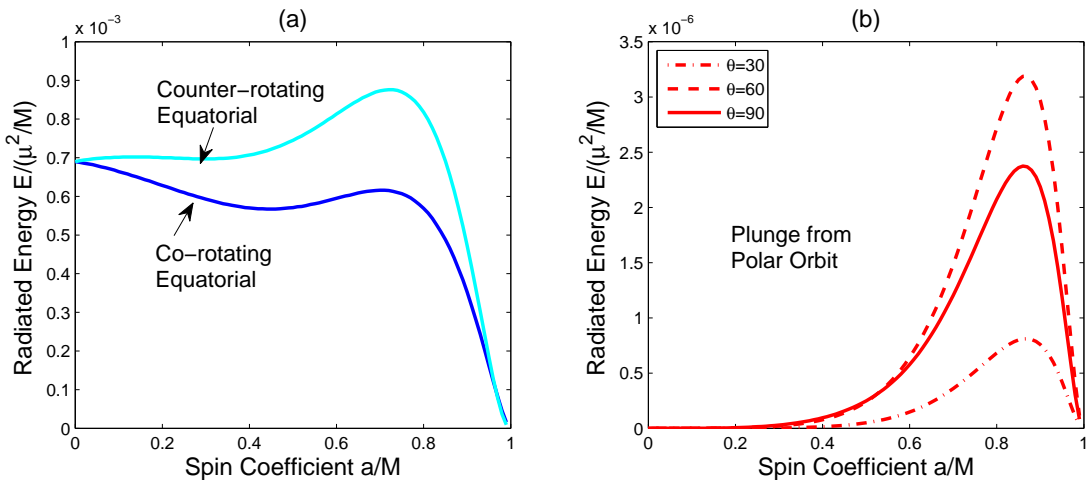}
\caption{Total energy radiated during plunging phase of orbit from ISCO as a function of spin parameter. Plot (a) displays the energy radiated when plunging from an equatorial ISCO (C=0) in the co- and counter-rotating cases.  Plot (b) shows the radiated energy if the particle originates from a polar orbit for various horizon impact angles, $\theta ={\pi \over 6},{\pi \over 3},{\pi \over 2} $ }
\label{pic:PLungeE}
\end{figure*}

Fig.\ref{pic:PLungeE} displays the total energy radiated during the plunging stage of orbits originating from ISCO. The factor $e^{-(\kappa/r_+)(T-t_0)}$ present in Eq.(\ref{eq:rad_E}) is weakly dependent on the orbital trajectory and set to unity for simplicity. Fig.\ref{pic:PLungeE} plot (a)  shows the energy radiated if the original orbit is equatorial. The non-zero $z$-component of the angular momentum ($L_z$) results in finite energy being radiated in the limit of zero spin, i.e. $a/M=0$. As the background spin increases from zero, 
the radiated energy of the co-rotating plunging orbit decreases. This effect is due to an effective decrease in the angular momentum resulting from the frame-dragging of the background black hole -- the factor of $(aE-L_z)^4$ in Eq.(\ref{eq:rad_E}) makes this apparent. 
In  the counter-rotating case, the same effect results in a net increase in energy radiated as spin increases from zero.
As the spin further increases, the competition between the increasing quality factor associated with the radiation and the suppression effect due to the effective potential barrier, described in the previous paragraph, sets in. A maximum in radiated energy is achieved at  $a/M=0.7$ and $a/M=0.72$ for the co- and counter-rotating cases respectively. 

Fig.\ref{pic:PLungeE} plot (b) shows the energy radiated from the end of the plunging phase when the particle plunges from a polar ISCO orbit. The azimuthal angular momentum ($L_z$) vanishes for this case and the resultant energy radiated approaches zero in the limit of zero spin,  $a/M=0$. 
As we increase the spin of the background black hole, 
an effective angular momentum represented 
by the factor $(aE-L_z)^4$ in Eq.(\ref{eq:rad_E}) comes into play 
and the radiated energy increases. For particles plunging from a polar orbit, the radiation suppression effect  noted in the equatorial case is also observed for high spin factors and a maximum in radiated energy is reached. This peak occurs at  a spin factor of $a/M=0.87$ 
when the impact angle with the horizon is $\theta_0=\pi/6$ or $\theta_0=\pi/3$ 
and at $a/M=0.86$ when the impact angle is $\theta_0=\pi/2$.  The best fit to the slopes of the three energy profiles in Fig.\ref{pic:PLungeE} plot (b) for the range of  spin factor between $0.7< a/M <0.8$  is found to be  $\Delta E = 3.22 \times 10^{-6} (a/M) -18.94 \times 10^{-6}$ 
for $\theta_0 = \pi/6$, 
$\Delta E = 1.24 \times 10^{-5} (a/M) -7.20 \times 10^{-6}$ 
for $\theta_0 = \pi/3$ 
and 
$\Delta E = 0.87 \times 10^{-5} (a/M) -5.89 \times 10^{-6}$ 
for $\theta_0 = \pi/3$ .

\begin{figure}[h]
\epsfxsize=\columnwidth 
\epsfbox{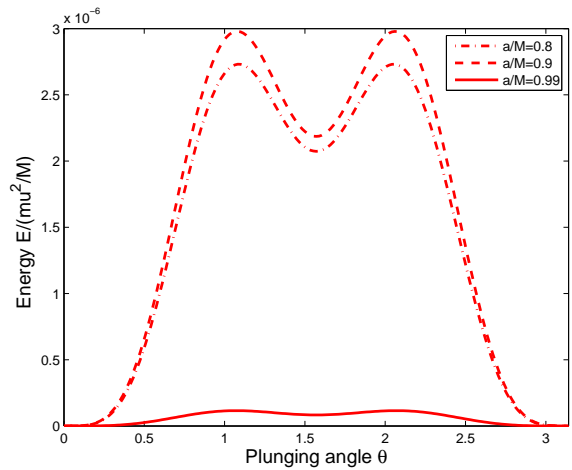}
\caption{Total energy radiated during   plunging phase  of a particle originating on a polar ISCO as a function of impact angle with the horizon. Radiation for background Kerr black holes with spin parameter $(a/M)=0.8,0.9,0.99$ are plotted.}
\label{fig:E3x}
\end{figure}

In the case of a particle plunging from a polar orbit, the radiated energy is dependent on the angle made with the spin axis when the particle impacts the horizon. The angular dependence of the radiated energy is displayed in Fig.\ref{fig:E3x} for various values of spin parameter.

Observe that the maximum in radiated energy 
does not occur when the particle falls in on the equatorial plane, 
$\theta_0=\pi/2$, but that it occurs instead for impact angles of $\theta_0=1.07$ and $\theta_0=2.06$. One would naively expect to find that the radiated energy is largest in the configuration where the particle has  maximum orbital velocity $v$ when it impacts the horizon, namely on the equatorial plane at  $\theta_0=\pi/2$. An initial scaling argument would be that the energy radiated is roughly proportional to the square of the metric perturbation  which  in turn is proportional to the square of the stress energy tensor of the source term $T\propto v^2 \propto (\omega r)^2 \sin^2 \theta_0$. This effect is indeed displayed by the appearance of  the factor of $\sin^4\theta_0$ in Eq.(\ref{eq:rad_E}). The  energy emission is, however, complicated by the angular pattern resulting from the spin-$2$ nature of gravitational waves, which is described by spheroidal harmonics  that enter Eq.(\ref{eq:rad_E}).
The origin of the double maxima in the emitted energy can be explained by focusing on the $(\ell m)$-sum of Eq.(\ref{eq:rad_E0}). Making use of equation  (\ref{eq:btran}), to evaluate the ratio  $B^{trans}$/ $B^{in}$, the sum can be expressed as 
\begin{eqnarray}
\sum_{\ell m}|\bar\omega_m|^6 |S_{\ell m \bar\omega_m}|^2 
\,, 
\end{eqnarray}
with $Re[\bar\omega_m] = m\Omega_H$. The dominant contribution to the  sum in Eq.(\ref{eq:rad_E0}) and  thus the  radiated energy is  made by modes with the largest $|m|$ value for a given $\ell$.  
If we consider the contribution from $(\ell=2)$-modes 
using the low-frequency result in App.\ref{app:teu1}, 
the $\theta_0$-dependence becomes  
\begin{eqnarray}
\sin^4(\theta_0) \left\{
\left(1-\cos(\theta_0)\right)^4+
\left(1+\cos(\theta_0)\right)^4\right\} 
\,, \nonumber 
\end{eqnarray}
which has two maxima at $\theta_0=1.06$ and $\theta_0=2.08$ 
and a local minima at $\theta_0=\pi/2$, and thus  accurately explains the $\theta_0$-dependence observed in Fig.\ref{fig:E3x}.

\section{Linear Momentum Flux} 
\label{sec:mom}
During the evolution of a typical binary system, the longest period of time is spent in the inspiral phase.  It is during this phase that most of the gravitational energy is emitted. By comparison to the inspiral phase the energy flux, computed in Sec \ref{sec:ene},  emitted during the plunging phase is insignificant. The situation is very different when one considers the linear momentum flux. The net linear momentum emission during the inspiral phase is small, because the binary orbit gradually progresses through a series of quasi-stable circular orbits resulting in nearly isotropic momentum emission.  As a result, the momentum emitted during the plunging stage is important and may be comparable to that emitted during inspiralling phase \cite{kickBH}. In this section we quantify the linear momentum flux carried away by gravitational waves during the final plunge phase of a spinning black hole.

The emitted linear momentum flux is given by 
\begin{align}
{dP_i \over dt} &= \lim_{r \to \infty} 
\left[{r^2 \over 16\pi}\int d\Omega \   n_i
\left| \int^t_{-\infty} dt \Psi_4 \right|^2 \right] 
\nonumber \\ 
&\to {\pi^2 \over 8} e^{-{\kappa \over r_+}(t-t_0-r^*)}
\sum_{\ell m,\ell' m'} 
{e^{-i(m-m')\Omega_H(t-r^*)} \over \bar\omega_m (\bar\omega_{m'})^*} \notag\\ &
\times {1 \over 2\pi}\int d\Omega \  n_i \  e^{i(m-m')\phi}
S_{\ell m \bar\omega_m}(S_{\ell' m' \bar\omega_{m'}})^*\notag\\
& \qquad  \qquad\times \tilde Z_{\ell m \bar\omega_m}
(\tilde Z_{\ell' m' \bar\omega_{m'}})^*
\,. \label{eq:rad_P0} 
\end{align}
where $n_i = (\sin\theta \cos\phi, \sin\theta \sin\phi, \cos\theta) 
$.

The $z$-component of the momentum flux, $dP_z/dt$, vanishes near the horizon.
This is yet another manifestation of the phenomenon discussed in Sec.\ref{sec:geo}, where the particle is frozen onto the horizon due to the time dilation effect and its motion is dominated by the velocity component in the $\phi$-direction. 
In the near horizon expansion of the  source term given in App.\ref{app:Source}, at the leading order only, the particle motion in the  $x$- and $y$-directions are taken into account and the net emission of $z$-linear momentum vanishes as a result.

\begin{widetext}
The  calculated  total linear momentum 
carried away by gravitational waves 
during the period $T <t-r^*< \infty$ is 
\begin{align}
\Delta (P_x+iP_y)
&= {\pi \over 16} \mu^2 {\kappa^5 M^2 \over r_+} 
{e^{-{\kappa \over r_+}(T-t_0)-i\Omega_H T} 
\over 1-i{a \over 2\kappa M}}
{\left(aE-{L_z\over\sin^2\theta_0}\right)^4 
\over (E-\Omega_H L_z)^2} \sin^4\theta_0 
\sum_{\ell \ell' m}
{1 \over \bar\omega_m^2((\bar\omega_{m'})^*)^2} 
S_{\ell m \bar\omega_m}(\theta_0)
\left(S_{\ell' m' \bar\omega_{m'}}(\theta_0)\right)^*
\notag\\
& \qquad \qquad \times \left({B^{trans}_{\ell m \bar\omega_m}
\over B^{in}_{\ell m \bar\omega_m}}\right)
\left({B^{trans}_{\ell' m' \bar\omega_{m'}}
\over B^{in}_{\ell' m' \bar\omega_{m'}}}\right)^* 
\int^\pi_0 d\theta \sin^2(\theta)
S_{\ell m \bar\omega_m}(\theta)
\left(S_{\ell' m' \bar\omega_{m'}}(\theta)\right)^* \,, 
\nonumber 
\end{align}
where we set  $m'=m+1$. 
To facilitate explicit evaluation, 
 the analytic expressions 
for the Teukolsky functions in the low frequency 
given in App.\ref{app:teu1} are used. 
The resulting expression is
\begin{align}\Delta (P_x+iP_y)
&\approx {\pi \over 1024} \mu^2 
{\kappa^{11} M^2 \over r_+^7} 
{e^{-{\kappa \over r_+}(T-t_0)-i\Omega_H T +i{a \over 2\kappa M}\ln\kappa} 
\over 1-i{a \over 2\kappa M}}
{\left(aE-{L_z\over\sin^2\theta_0}\right)^4 
\over (E-\Omega_H L_z)^2} \sin^5\theta_0 
\nonumber \\ & \times 
\Biggl\{{16 \over 135}
-i{16 \over 45}\left({a \over \kappa M}\right)
+\left({16 \over 45}-{8 \over 15}\sin^2(\theta_0)\right)
\left({a \over \kappa M}\right)^2
\nonumber \\ & \qquad 
+i\left(-{176 \over 135}+{16 \over 15}\sin^2(\theta_0)\right)
\left({a \over \kappa M}\right)^3
+\left({32 \over 45}-{8 \over 15}\sin^2(\theta_0)\right)
\left({a \over \kappa M}\right)^4
\nonumber \\ & \qquad 
+i\left(-{64 \over 45}+{16 \over 15}\sin^2(\theta_0)\right)
\left({a \over \kappa M}\right)^5
+\left({128 \over 135}-{32 \over 45}\sin^2(\theta_0)\right)
\left({a \over \kappa M}\right)^6
\Biggr\}
\,. \label{eq:rad_P} 
\end{align}

\begin{figure*}[t]
\includegraphics[width=\columnwidth]{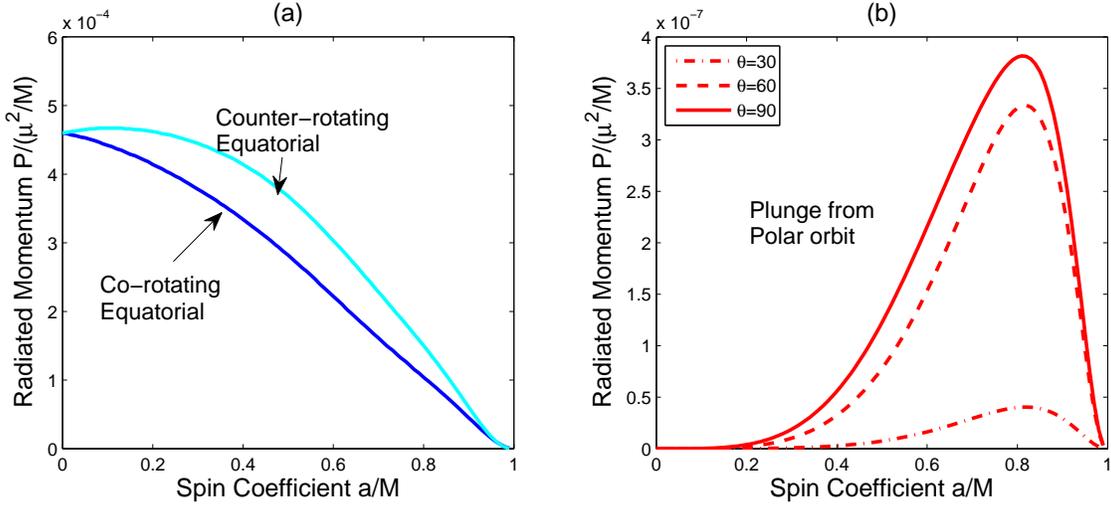}
\caption{Total momentum radiated during the final plunging phase of orbit at ISCO as a function of spin parameter. Plot (a) displays the momentum radiated when plunging from an equatorial ISCO ($C=0$) in the co- and counter-rotating cases.  Plot (b) shows the radiated momentum if the particle originates from a polar orbit  ($L_z=0$)  for various horizon impact angles, $\theta ={\pi \over 6},{\pi \over 3},{\pi \over 2} $  }
\label{fig:P12}
\end{figure*}
\end{widetext}

The momentum radiated during the final plunging phase of an orbit originating at ISCO on the equatorial plane is plotted in Fig \ref{fig:P12} (a). 
Analogous  to the radiated energy of the same orbit, Fig \ref{pic:PLungeE} (a),
the linear momentum plotted here displays a finite value in the limit of a non-spinning Kerr background hole due to the orbital angular momentum of the plunging particle.
The momentum radiation further displays features reminiscent of the radiated energy, in that the radiated momentum decreases for the co-rotating plunge and displays an increase at first followed by a decrease for the counter-rotating plunge
as the  background spin increases from zero.
A feature distinct from those displayed by the  radiated energy in Fig. \ref{pic:PLungeE} (a) is that no local maxima of the radiated momentum is apparent.  This is due to the factor $1/(1-i(a/2\kappa M))$ 
which comes from the time integration of the momentum flux.  
This can be better understood by observing that in Eq.(\ref{eq:rad_P0}), 
the momentum flux displays damped oscillatory behavior and for large spin, the oscillation frequency becomes high. When taking the integral to obtain the total radiated momentum, the positive and negative parts of the integrand cancel, resulting in a small value for the integral.  This stands in contrast to the radiated energy where the integrand is a positive decaying function without any oscillatory behavior.

The oscillatory behavior of the momentum flux  made explicit in Eq.(\ref{eq:rad_P0}) indicates 
that the linear momentum flux is emitted in various 
directions in x/y-plane. This feature is to be expected if the particle is in effect pulled into orbit around the black hole as it plunges and so the radiated momentum has an oscillatory nature that will  reflect  the horizon velocity of the final black hole.
A similar process was alluded to  in Fig.14 in the second reference of \cite{kickSPe} to explain the anti-kick phenomena observed in \cite{akick}.

The momentum radiated during the final stages of plunge from a polar ISCO is shown in In Fig.\ref{fig:P12} (b).
Analogous to the radiated energy plotted in Fig.\ref{pic:PLungeE} (b), 
the radiated momentum increases from zero 
as the background spin increases. The radiated momentum reaches a  maximum at $a/M=0.82$,  when the in-falling angle is $\theta_0=\pi/6$ or $\theta_0=\pi/3$ 
and at $a/M=0.81$  when the in-falling angle is $\theta_0=\pi/2$. 
As the spin-factor ranges over $0.55< a/M <0.75$, the radiated momentum can be fitted by  
$\Delta P = 1.23 \times 10^{-7} (a/M) -5.65 \times 10^{-8}$ 
for $\theta_0 = \pi/6$ 
$\Delta P = 0.95 \times 10^{-6} (a/M) -4.12 \times 10^{-7}$ 
for $\theta_0 = \pi/3$ 
and 
$\Delta P = 0.94 \times 10^{-6} (a/M) -3.62 \times 10^{-7}$ 
for $\theta_0 = \pi/3$. 

\begin{figure}[h]
\includegraphics[width=\columnwidth]{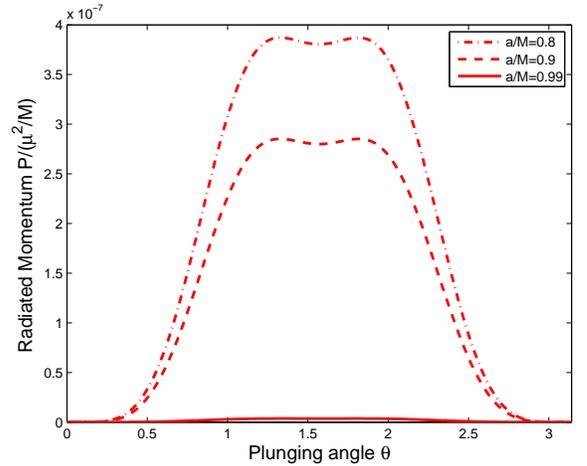}
\caption{ Total momentum  radiated during plunging phase  of a particle originating on a polar ISCO as a function of impact angle with the horizon. Radiation for background Kerr black holes with spin parameter $(a/M)=0.8,0.9,0.99$ are plotted.}
\label{fig:P3x}
\end{figure}

The dependence of the radiated momentum on the angular position at which the particle impacts the horizon during a plunge from a polar orbit is show in In Fig.\ref{fig:P3x}.
The angular profiles of the radiated momentum (Fig.\ref{fig:P3x}) 
and those of the radiated energy (Fig.\ref{fig:E3x}) differ slightly. 
The features of the local minimum at $\theta_0=\pi/2$ 
and the two maxima off the equatorial plane are robust. For the case of radiated momenta, the maxima occur at $\theta_0=1.33$ and $\theta_0=1.82$. 
Following an argument similar to that employed in the case of the radiated energy, the observed behavior can be explained by analyzing 
the $\theta_0$-dependence of the $(\ell \ell' m)$-sum 
in the first line of equation (\ref{eq:rad_P}). 
The dominant functional behavior is found to be
\begin{eqnarray}
\sin^5(\theta_0) \left\{
\left(1-\cos(\theta_0)\right)^3+
\left(1+\cos(\theta_0)\right)^3\right\} 
\,, \nonumber 
\end{eqnarray}
which admits maxima at $\theta_0=1.35$ and $\theta_0=1.79$ 
and a local minima at $\theta_0=\pi/2$.

\section{conclusion} \label{sec:con}

This paper explores the dynamics of 
the final stages of the plunge
of a small black hole into a Kerr black hole and characterizes the resulting gravitational radiation, using perturbation theory. The purpose  of this exploration is to gain analytic understanding of the features which dominate the dynamics of this transition phase of the merger process before the onset of quasi-normal ringing and to identify  any universal features. Our analysis strives to highlight the effects of the spin of the central black hole on  the resulting gravitational waveform and momentum flux while building an intuitive picture of the processes responsible for their origin. It should be noted that the all graphs plotted in this paper and comments made are relevant only to radiation emanating from the so-called near horizon region during the plunging phase.

In Sec.\ref{sec:mp}, it was found that the waveform resulting from the plunging phase can be expressed as 
\begin{align}
h(t) &= \sum_{n, m=1,2 \cdots} 
h_{(m,n)}e^{-i\omega_{(m,n)}t} 
\, , \label{waveT}
\end{align}
where the frequencies entering the expansion take on the universal  form $\omega_{(m,n)} = m\Omega_H t -in\kappa /( 2 r_+)$. Therefore, the time dependence of the emitted radiation is entirely determined by the characteristics of the massive central black hole.  This waveform describes  the transition period between radiation dominated by the inspiral and radiation dominated by the free quasi-normal ringing.  The inspiral stage broadcasts the characteristic frequencies of the orbiting particle and is strongly affected by the constants of motion describing the orbit. The quasi-normal ringing phase is characterized by the frequencies and damping rates associated with free vacuum perturbations of an excited Kerr black hole as it settles down to  equilibrium. 
The frequencies that enter expression \eqref{waveT} can be ascribed to neither of the two preceding scenarios.  They describe a state of forced oscillation distinct in frequency from free quasi-normal ringing but are at the same time devoid of any characteristic identifying the source of the perturbation.  In fact, the source  can be best described as a reluctant mirage on the horizon -- a mirage dragged into motion and then retained for distant observation by an all-dominating frame dragging effect. The frequency associated with $\omega_{(m,n)}$ originates from the azimuthal motion of the orbit on the horizon, which is entirely determined by parameters of the background black hole.
 The method by which this mirage is stripped of its identity while it in essence passes through the event horizon and is assimilated into the large black hole is described in  mathematical detail in Sec.\ref{sec:geo}.

The wave amplitude $h_{(m,n)}$ still retains characteristics of the source and depends on the orbital constants.
 The damping rate associated with $\omega_{(m,n)}$  indicates that the radiation corresponds to a decaying part of the waveform. A feature that this frequency shares with the quasi-normal mode frequencies is the existence of a high quality factor in the maximally spinning limit $a/M\to 1$, for example,  $Q=14.04$ for $a/M=0.99$ 
and $Q=141.41$ for $a/M=0.99$. A more detailed comparison between the least damped quasi-normal mode frequency and $\omega_{(2,1)}$ is given toward the end of Sec.\ref{sec:mp}.

The gravitational flux of energy and linear momentum originating from the plunging phase is computed in Sec.\ref{sec:ene} and Sec.\ref{sec:mom}, respectively. Orbits plunging from ISCO in the equatorial plain (i.e. $C=0$) and from a polar ISCO (i.e. $L_z=0$) have been considered. In general, the following characteristics can be identified.  The rapid increase in quality factor 
is counteracted by a term in the amplitude  $h_{(m,n)}$ that rapidly approaches zero as spin approaches maximality. This  ensures that all forms of radiation approach zero in the limit of maximal spin and that the total emitted energy and momentum during this phase remain finite. This effect further ensures that a maximum in emitted energy as a function of spin is attained in all cases considered. For  the plunge from an equatorial orbit the maximum in radiated energy occurs around  $a/M=0.7$  and  for the plunge from polar orbit the maximum occurs for a spin parameter around $a/M=0.86$.

The plunge from a  polar ISCO orbit is of particular interest because it can be used as a toy problem to illuminate 
some of the features observed in numerical simulations of black hole mergers in super-kick configurations. In particular, the dependence 
of the emitted linear momentum on the  angle $\theta_0$ made with the rotation axis when the particle enters the horizon is computed in this paper.  Contrary to expectation, it was found in Sec.\ref{sec:mom} that the produced profile does not have a simple sinusoidal dependence on angle, but rather that it is complicated by the spin-2 nature of gravitational radiation, as displayed in Fig.\ref{fig:P3x}.  

The radiated linear momentum originating from a polar plunge reaches a maximum with respect to the spin parameter at around $a/M=0.86$. While the origin of this peak is well understood within  the context of our model, the question remains as to, whether the peak is an artifact of linear perturbation theory and our underlying assumptions  or  whether one will find its counterpart in numerical simulations. We also find that the radiated energy and linear momentum 
can be fitted by linear functions of $a/M$ 
in the domain $0.7< a/M <0.8$ and $0.55< a/M <0.75$, 
respectively. 

A final observation regarding  the computed momentum flux of Sec.\ref{sec:mom}  Eq.(\ref{eq:rad_P0}) is the existence of  oscillatory behavior with a frequency we now consider characteristic of a  plunging event involving spin.  The introduced phase implies that the linear momentum flux is emitted in various directions in the x/y-plane, consistent with our picture of a miraged particle stuck on the event horizon radiating momentum in a manner that reflects the velocity of the final black hole while it slowly recedes from view.  This could provide a heuristic explanation of the origin of the anti-kick phenomenon observed in \cite{akick} and is suggestive that the anti-kick is indicative of the final merger event.


\section{Acknowledgment}
We would like to thank Prof. Kip Thorne, Prof. Yanbei Chen 
and Dr. Emanuele Berti for fruitful discussions. 
YM is supported 
by NSF grant PHY-0601459, PHY-0653653, 
NASA grant NNX07AH06G, NNG04GK98G 
and the Brinson Foundation. JB gratefully acknowledges support from the Sherman Fairchild Prize Postdoctoral Fellowship Program. 


\appendix 

\section{Teukolsky formalism} \label{app:teu0}
This appendix summarizes the Teukolsky formalism used in the derivation of the Weyl scalar $\Psi_4$  which contains the radiation content of the perturbed spacetime with a prescribed source term.

The  scalar $\Psi_4$ can be decomposed into its  Fourier-harmonic components as follows
\begin{align}
\Psi_4 &=& {1 \over (r-ia\cos\theta)^4} 
\sum_{\ell m} \int d \omega  e^{-i\omega t +im\phi} 
R_{\ell m\omega}(r) S_{\ell m\omega}(\theta) 
\,. \label{eq:cuv} 
\end{align}
where $ S_{\ell m\omega}(\theta) $ and $R_{\ell m\omega}(r)$  represent the angular and radial Teukolsky functions respectively.
 The angular function, obeys the angular Teukolsky equation for $s=-2$,  explicitly,
\begin{align}
{1\over \sin\theta}{d \over d\theta}
\left(\sin\theta{dS_{\ell m\omega} \over d\theta}\right)
-US_{\ell m\omega} &= 0 \, , \label{eq:teuAng} \end{align}
where the potential $U$ is given by
\begin{align} 
U= a^2\omega^2\sin^2\theta 
+{(m-2\cos\theta)^2\over \sin^2\theta} \notag\\
\qquad -4a\omega\cos\theta+2-2ma\omega-\lambda 
\,, 
\end{align}
and $\lambda$ denotes the eigenvalue of $S_{\ell m\omega}$. 
The angular Teukolsky function is normalized so that,  
\begin{align}
\int^\pi_0 \sin\theta\  d\theta |S_{\ell m\omega}|^2 =1 
\,. 
\end{align}
The radial Teukolsky function in turn satisfies 
\begin{align}
\Delta^2{d \over dr}\left({1\over \Delta}
{dR_{\ell m\omega}\over dr}\right)
-VR_{\ell m\omega} &= T_{\ell m\omega} 
\,, \label{eq:teu} 
\end{align}
where the potential $V$ is defined to be
\begin{align}
V &= -{K^2+4i(r-M)K \over \Delta} +8i\omega r +\lambda 
\,
\end{align}
and the function $K = (r^2+a^2)\omega -ma$. The source term $T_{\ell m\omega}$ contains the effect of the stress-energy tensor perturbing the background spacetime. The relationship  between the source term and the stress-energy tensor is given by
\begin{align}
T_{\ell m\omega} &= 4\int d\Omega dt 
\rho^{-5} \bar\rho^{-1}(T_1+T_2)e^{-im\phi+i\omega t}
{S_{\ell m\omega} \over \sqrt{2\pi}} 
\,, \\ 
T_1 &= 
-{1 \over 2}\rho^8\bar\rho{\cal L}_{-1}
[\rho^{-4}{\cal L}_0(\rho^{-2}\bar\rho^{-1}T_{nn})]
\nonumber \\ &
-{1 \over 2\sqrt{2}}\rho^8\bar\rho\Delta^2{\cal L}_{-1}
[\rho^{-4}\bar\rho^2{\cal D}_+(\rho^{-2}\bar\rho^{-2}
\Delta^{-1}T_{\bar mn})]
\,, \\ 
T_2 &=
-{1 \over 4}\rho^8\bar\rho\Delta^2{\cal D}_+
[\rho^{-4}{\cal D}_+(\rho^{-2}\bar\rho T_{\bar m\bar m})]
\nonumber \\ &
-{1 \over 2\sqrt{2}}\rho^8\bar\rho\Delta^2{\cal D}_+
[\rho^{-4}\bar\rho^2\Delta^{-1}{\cal L}_{-1}
(\rho^{-2}\bar\rho^{-2}T_{\bar mn})]
\,, 
\end{align}
where we use 
$\rho = (r-ia\cos\theta)^{-1} 
\,$, 
 $\bar\rho = (r+ia\cos\theta)^{-1} 
\,,$ and the operators 
$
{\cal D}_+ = \partial_r+iK/\Delta 
\,,$ and $
{\cal L}_s= \partial_\theta +{m\over \sin\theta}
-a\omega \sin\theta +s \cot\theta
$. The terms $T_{nn}$, $T_{\bar mn}$ and $T_{\bar m \bar m}$ 
denote the tetrad components of the stress-energy tensor. 
In this paper, a point particle was used  
as the source of the metric perturbation. The stress-energy tensor associated with a point particle can be expressed as 
\begin{align}
T^{\mu\nu} = &{\mu \over \Sigma \sin\theta}
\left({dt \over d\tau}\right)^{-1}
{dz^\mu \over d\tau}{dz^\nu \over d\tau}\notag\\
&\qquad  \times \delta(r-r(t))\delta(\theta-\theta(t))\delta(\phi-\phi(t))
\,, 
\end{align}
where $\mu$ is the mass of the particle. 
Using the geodesic equations (\ref{eq:geo_r})-(\ref{eq:geo_phi}), the source term can be simplified to,  
\begin{align}
T_{lm\omega} &= \mu \int dt e^{i\omega t -im\phi}
\Delta^2\bigl[\partial_r^2\left\{A_{\bar m\bar m2}\delta(r-r(t))\right\}
\nonumber \\ & \qquad\qquad
+\partial_r\left\{(A_{\bar mn1}+A_{\bar m\bar m1})
\delta(r-r(t))\right\}
\nonumber \\ & \qquad\qquad
+(A_{nn0}+A_{\bar mn0}+A_{\bar m\bar m0})\delta(r-r(t))
\bigr]
\,, \end{align}
where
\begin{align}
A_{nn0} &= {-2 \over \sqrt{2\pi}\Delta^2}
B_{nn} \rho^{-2}\bar\rho^{-1}{\cal L}^\dagger_1
\left\{\rho^{-4}{\cal L}^\dagger_2(\rho^3 S)\right\}
\,, \\ 
A_{\bar mn0} &= {2 \over \sqrt{\pi}\Delta}
B_{\bar mn} \rho^{-3}\left[
\left({\cal L}^\dagger_2 S\right)
\left({iK \over \Delta}+\rho+\bar\rho\right))\right.\notag
\\
& \qquad \qquad \ \ \ \  \  \ \ \ \ \ \ \ \ \left .
-a\sin\theta S{K\over \Delta}(\bar\rho-\rho)
\right]
\,, \\ 
A_{\bar m\bar m0} &= -{1 \over \sqrt{2\pi}}
\rho^{-3}\bar\rho B_{\bar m\bar m}S\left[
-i\left({K \over \Delta}\right)_{,r}
-{K^2 \over \Delta^2}+2i\rho{K \over \Delta}
\right]
\,, \\ 
A_{\bar mn1} &= {2 \over \sqrt{\pi}\Delta}
\rho^{-3}B_{\bar mn}\left[
{\cal L}^\dagger_2 S +ia\sin\theta S(\bar \rho-\rho)
\right]
\,, \\ 
A_{\bar m\bar m1} &= -{2 \over \sqrt{2\pi}}
\rho^{-3}\bar\rho B_{\bar m\bar m}S\left(
i{K \over \Delta}+\rho\right)
\,, \\ 
A_{\bar m\bar m2} &= -{1 \over \sqrt{2\pi}}
\rho^{-3}\bar\rho B_{\bar m\bar m}S
\,, 
\end{align}
and
\begin{align} 
B_{nn} &= {1\over 4\Sigma^3 \dot t}
\left[E(r^2+a^2)-aL_z+\Sigma{dr \over d\tau}\right]^2 
\,, \\ 
B_{\bar mn} &= -{\rho\over 2\sqrt{2}\Sigma^2 \dot t}
\left[E(r^2+a^2)-aL_z+\Sigma{dr \over d\tau}\right] \notag\\
&\qquad \qquad \ \ \ \ \ \times
\left[i\sin\theta\left(aE-{L_z \over \sin^2\theta}\right)\right] 
\,, \\ 
B_{\bar m\bar m} &= {\rho^2\over 2\Sigma \dot t}
\left[i\sin\theta\left(aE-{L_z \over \sin^2\theta}\right)\right]^2 
\,. 
\end{align}
Here the operator ${\cal L}^\dagger_s$  is defined by $
{\cal L}^\dagger_s = \partial_\theta -{m\over \sin\theta}
+a\omega \sin\theta +s \cot\theta
\, $
, indices have been dropped from the angular Teukolsky function, i.e. $S=S_{\ell m\omega}$ 
and $\dot t=dt/d\tau=(1/\Sigma)(dt/ds)$. 

For a given source the  particular solution of the radial Teukolsky equation can be found by means of a Green's function method.  To this end, define two homogeneous solutions 
to the radial Teukolsky equation which satisfy the boundary conditions 
\begin{align}
R^{in}_{\ell m\omega} &\to \left\{\begin{matrix}
B^{trans}_{\ell m\omega}\Delta^2 e^{-ikr^*} 
& \hbox{for $r\to r_+$} \cr 
B^{ref}_{\ell m\omega} r^3e^{i\omega r^*} 
+B^{in}_{\ell m\omega} r^{-1}e^{-i\omega r^*} 
& \hbox{for $r\to\infty$} 
\end{matrix} \right. 
\,, \label{eq:in_r_teu} \\ 
R^{up}_{\ell m\omega} &\to \left\{\begin{matrix}
C^{up}_{\ell m\omega} e^{ikr^*} 
+C^{ref}_{\ell m\omega}\Delta^2 e^{-ikr^*} 
& \hbox{for $r\to r_+$} \cr 
C^{trans}_{\ell m\omega} r^3e^{i\omega r^*} 
& \hbox{for $r\to\infty$} 
\end{matrix}\right. 
\,, \label{eq:up_r_teu} 
\end{align}
in these expressions  $k=\omega-ma/2Mr_+$ 
and $r^*$ is the tortoise coordinate defined by 
$(dr^*/dr)=(r^2+a^2)/\Delta$. 
In terms of homogeneous solutions the particular solution of the  radial Teukolsky function can be expressed as  
\begin{align}
R_{\ell m\omega} &= {1  \over W_{\ell m\omega}}
R^{up}_{\ell m\omega} \int^r_{r_+}dr 
R^{in}_{\ell m\omega} \Delta^{-2}T_{\ell m\omega}\notag\\
&
\ \ \ \ + {1  \over W_{\ell m\omega}} R^{in}_{\ell m\omega} \int^\infty_r dr 
R^{up}_{\ell m\omega} \Delta^{-2}T_{\ell m\omega}
\,, 
\end{align}
where the Wronskian $W_{\ell m\omega}$   is given by \mbox{$
W_{\ell m\omega} = 2i\omega 
B^{in}_{\ell m\omega}C^{trans}_{\ell m\omega} 
\,. $}
Of particular interest for the purposes of gravitational wave extraction is the nature of the solution at infinity,
\begin{align}
R_{\ell m\omega}(r\to\infty) &\to 
{r^3 e^{i\omega r^*} \over 2i\omega B^{in}_{\ell m\omega}} 
\int^\infty_{r_+} dr 
R^{in}_{\ell m\omega} \Delta^{-2}T_{\ell m\omega} \notag\\
& \qquad \qquad \qquad = r^3 e^{i\omega r^*} Z_{\ell m\omega} 
\,. \label{eq:teuk_r} 
\end{align}
In the event that the source term 
is a point particle moving along a geodesic, such as the scenario considered in this paper, 
the function $Z_{\ell m\omega}$ can be expressed as 
\begin{widetext}
\begin{align}
Z_{\ell m\omega} &=
{\mu \over 2i\omega B^{in}_{\ell m\omega}}
\int dt e^{i\omega t-im\phi(t)} \bigl[X_{\ell m\omega}
+X_{\ell m\omega \infty}-X_{\ell m\omega +}\bigl] 
\,, 
\end{align}
where \begin{align}
X_{\ell m\omega} &=& \left[
R^{in}_{\ell m\omega}
\left\{A_{nn0}+A_{\bar mn0}+A_{\bar m\bar m0}\right\}
-{dR^{in}_{\ell m\omega} \over dr}
\left\{A_{\bar mn1}+A_{\bar m\bar m1}\right\}
+{d^2R^{in}_{\ell m\omega} \over dr^2}
A_{\bar m\bar m2}
\right]_{r=r(t)} 
\,, \label{eq:xx0} \\ 
X_{\ell m\omega \infty} &=& 
\lim_{r(t)\to \infty}e^{i\omega t-im\phi(t)}
\left[R^{in}_{\ell m\omega}
\left\{A_{\bar mn1}+A_{\bar m\bar m1}\right\}
-\left\{i\left(\omega{dt \over dr}-m{d\phi\over dr}\right)
R^{in}_{\ell m\omega}
+2{dR^{in}_{\ell m\omega} \over dr}\right\}
A_{\bar m\bar m2}
\right]{dt \over dr}
\,, \label{eq:xx1} \\ 
X_{\ell m\omega +} &=& 
\lim_{r(t)\to r_+}e^{i\omega t-im\phi(t)}
\left[R^{in}_{\ell m\omega}
\left\{A_{\bar mn1}+A_{\bar m\bar m1}\right\}
-\left\{i\left(\omega{dt \over dr}-m{d\phi\over dr}\right)
R^{in}_{\ell m\omega}
+2{dR^{in}_{\ell m\omega} \over dr}\right\}
A_{\bar m\bar m2}
\right]{dt \over dr}
\,. \label{eq:xx2} 
\end{align} the terms 
(\ref{eq:xx1}) and (\ref{eq:xx2}) appear 
as the boundary terms when equation  (\ref{eq:teuk_r}) is integrated by parts.
If the source particle orbit  is bound within the radial domain, namely somewhere between  infinity and the horizon, 
these terms vanish.

For the source under consideration in the present paper,  care has to be exercised since the orbit crosses the horizon and the term (\ref{eq:xx2}) may contribute to the integral. It is however found that this term too vanishes, 
which is reasonable 
since  no waves propagate 
from the horizon out to infinity.

\section{Leading order behavior of Source Terms}
\label{app:Source}
The source terms generated by a particle plunging on a geodesic orbit near the horizon of a black hole display the following leading order behavior when expanded  with respect to distance from the horizon
\begin{align}
B_{nn} &\to O((r-r_+)^3) 
\,, \quad 
B_{\bar mn} \to O((r-r_+)^2)
\,, \\ 
B_{\bar m\bar m} &\to 
-{a^2\kappa E_{ISCO} \over 4Mr_+}
{\sin^2\theta_0 \over (r_+-ia\cos\theta_0)^2}
{r-r_+ \over r_+} +O((r-r_+)^2)
\,, \\ 
A_{nn0} &\to O(r-r_+) 
\,, \quad 
A_{\bar mn0} \to O((r-r_+)^0) 
\,, \\ 
A_{\bar m\bar m0} &\to 
-{\kappa a^2 E_{ISCO} \over 4\sqrt{2\pi}Mr_+^2}
\left\{i{kr_+\over\kappa}
+\left({kr_+\over\kappa}\right)^2\right\}
{r_+-ia\cos\theta_0 \over r_++ia\cos\theta_0}
\sin^2\theta_0 S
{1 \over r-r_+} 
+O((r-r+)^0) 
\,, \\ 
A_{\bar mn1} &\to O(r-r_+)
\,, \\ 
A_{\bar m\bar m1} &\to 
i{\kappa a^2 E_{ISCO} \over 2\sqrt{2\pi} Mr_+^2}
{kr_+\over\kappa}
{r_+-ia\cos\theta_0 \over r_++ia\cos\theta_0}
\sin^2\theta_0 S
+O(r-r+) 
\,, \\ 
A_{\bar m\bar m2} &= 
{\kappa a^2 E_{ISCO} \over 4\sqrt{2\pi}M r_+^2}
{r_+-ia\cos\theta_0 \over r_++ia\cos\theta_0}
\sin^2\theta_0 S
(r-r_+) 
+O((r-r+)^2) 
\,, 
\end{align}
Note that the dominant contribution of the particle's geodesic motion enters the source terms by means of $B_{\bar m\bar m}$ . This term dominates due to the rapid azimuthal velocity induced by the frame dragging effect.
 \end{widetext}

\section{Teukolsky Functions in the  low-frequency limit} 
\label{app:teu1}

This appendix summarizes the the analytic considerations required to compute the functional form of homogeneous Teukolsky functions in the low-frequency limit.
 
\noindent
\subsection{ Angular Teukolsky Functions} 
The solutions to the angular Teukolsky equation \label{eq:teuAng} are also known as spin-weighted angular spheroidal functions.  The  analytic structure of these functions was studied in \cite{s_teu}. 
By expanding the spin-weighted spheroidal functions in terms of Jacobi polynomials the angular Teukolsky equation can be reduced to three-term recurrence relations that are given by Eqs.(21-24) of \cite{s_teu}.
The solutions of the recurrence relations were shown, in Eq.(25) of  \cite{s_teu}, to converge very rapidly with respect to an expansion in frequency, i.e.  with respect to expansion parameter  $a \omega$.  
As a result, in the  low-frequency limit,   Jacobi polynomials accurately capture  the analytic behavior of the spin-weighted spheroidal harmonics.

To leading order in the low frequency limit, the angular Teukolsky functions can thus be well approximated by  spin-weighted spherical harmonics. It is this approximation that is adopted in this paper to aid computational simplicity. In particular we have,
\begin{align}
S_{\ell m\omega}(\theta) \to {}_{-2}Y_{lm}(\theta,0) 
\,, \quad 
\lambda \to (\ell-1)(\ell+2) 
\,, 
\end{align}
where $\ell \geq |m|$ and $\ell=2,3,\cdots$. 
The  low $\ell$,  spin-weighted spherical harmonics used in the computation are now given  
\begin{align}
{}_{-2}Y_{2 \pm 2}(\theta,0) &= 
{1 \over 8}\sqrt{5 \over \pi}(1\pm\cos\theta)^2
\,, \\ 
{}_{-2}Y_{2 \pm 1}(\theta,0) &= 
{1 \over 4}\sqrt{5 \over \pi}\sin\theta(1\pm\cos\theta)
\,, \\ 
{}_{-2}Y_{20}(\theta,0) &= 
{1 \over 4}\sqrt{15 \over 2\pi}\sin^2\theta
\,. 
\end{align}

\subsection{ Radial Teukolsky Functions}
The properties of the solutions to the radial  Teukolsky equation used to aid calculation are briefly summarized in this section. The analysis of the analytic structure of the   radial Teukolsky equation was first performed by Leaver \cite{r_teu0} and further developed by Mano et. al .\cite{r_teu1}, it is this reference to which the reader should refer  for a treatment  more detailed than the synopsis given here. 

 The approach adopted in the analysis is to expand the homogeneous solutions in terms of two different series expansions, each strongly convergent at a particular boundary condition.
Near the horizon  hypergeometric functions were used (refer to  Sec.2 of Ref.\cite{r_teu1}) to capture the behavior of in-going and outgoing waves and implement the boundary condition on the horizon. An expansion in Coulomb functions was then used to give an asymptotic expression at infinity, and to fix the boundary condition of no in-going waves   (refer to  Sec.3 of Ref.\cite{r_teu1}).  These two expansions where then matched in the interior to obtain a solution valid over the whole domain obeying the appropriate boundary conditions on both boundaries (refer to  Sec.4 of Ref.\cite{r_teu1}). 

Following an approach similar to that employed in the case of the angular Teukolsky function,
the radial Teukolsky equation can be reduced 
to  three-term recurrence relations valid for both expansions. Once again the solution to the three term recurrence relations 
shows very rapid convergence with respect to a low frequency expansion. In particular the expansion parameter $\epsilon =2 \omega M$ is introduced. 

In the  calculation performed in this paper, 
the ratio of the incoming wave amplitude, $B^{in}$ to  the amplitude of the transmitted wave, $B^{trans}$ as defined in Eq.(\ref{eq:in_r_teu}) is required. 
Using \cite{r_teu1} the ratio of these wave amplitudes can be obtained in terms of a power series expansion in $\epsilon$. The rapid convergence of the expansions employed in  \cite{r_teu1} with respect to $\epsilon$ allows  us to consider only the leading order term in the   expansion to make a quantitative estimate  of the required ratio.
The wave amplitude ratio so calculated is
\begin{align}
{B^{trans} \over B^{in}} &= 
i^{1-\ell}2^{\ell+3} \omega^5 
e^{-i{m \over 2}{q \over \kappa}\ln\kappa}
\left(\kappa\epsilon\right)^{\ell-2} \notag\\
& \qquad {(\ell+2)!(\ell-2)! \over (2\ell)!(2\ell+1)!}
{\Gamma\left(\ell+1+im{q \over \kappa}\right)
\over \Gamma\left(3+im{q \over \kappa}\right)} 
\left(1 +O(\epsilon)\right) 
\ \label{eq:btran}, 
\end{align}
where $q=a/M$ and $\kappa = \sqrt{1-q^2}$. The factorials in the denominator of Eq. (\ref{eq:btran}), cause the computed ratio to decrease very rapidly as $\ell$ increases.  As a result only the lower  $\ell$ contributions need be considered.
Specifically, the lowest two terms contributing to the expression are computed to be
\begin{align}
\left[{B^{trans} \over B^{in}}\right]_{\ell=2} &= 
-i{4 \over 15}\omega^5 
e^{-i{m \over 2}{q \over \kappa}\ln\kappa}
\left(1 +O(\epsilon)\right) 
\,, \\ 
\left[{B^{trans} \over B^{in}}\right]_{\ell=3} &= 
-{2 \over 945}\omega^5 (\kappa\epsilon)
e^{-i{m \over 2}{q \over \kappa}\ln\kappa}
\left(3+im{q \over \kappa}\right)
\left(1 +O(\epsilon)\right) 
\,. 
\end{align}

\bibliographystyle{apsrev}

\begin{thebibliography}{99}


\bibitem{NR} 
Frans Pretorius, Relativistic Objects in Compact Binaries: From Birth to Coalescence,
Colpi et al., Springer Verlag, Canopus Publishing Limited,
arXiv:0710.1338, and references therein

\bibitem{skick1}
M. Campanelli, C. O. Lousto, Y. Zlochower, D. Merritt
Astrophys.J. 659 (2007) L5, 
J. A. Gonzalez, M. D. Hanna,, U. Sperhake, B. Brugmann and S. Husa, 
Phys. Rev. Lett. 98, 231101 (2007), 
M. Campanelli, C. O. Lousto, Y. Zlochower, D. Merritt
Phys. Rev. Lett 98, 231102 (2007). 

\bibitem{skick2}
J. G. Baker, W. D. Boggs, J. Centrella, B. J. Kelly, 
S. T. McWilliams, M. C. Miller, J. R. van Meter
Astrophys. J. 682, L29 (2008) 

\bibitem{skick3}
S. Dain, C. O. Lousto, Y. Zlochower
Phys. Rev. D 78, 024039 (2008)



\bibitem{kick}
J. D. Bekenstein J D 
Astrophys. J. Lett. 183, 657 (1973). 

\bibitem{kickPN}
M. J. Fitchett, 
Mon. Not. R. Astron. Soc. 203, 1049 (1983), 
A. G. Wiseman
Phys. Rev. D 46, 1517 - 1539 (1992). 

\bibitem{kickSP}
L. E. Kidder,
Phys. Rev. D 52, 821 (1995), 
S. H. Miller, R.A. Matzner
arXiv:0807.3028, to be publish at Gen. Rel. Grav. 

\bibitem{kickPNe}
L. Blanchet, M. S. S. Qusailah, C. M. Will
Astrophys.J. 635 (2005) 508, 
T. Damour, A. Gopakumar
Phys.Rev. D73 (2006) 124006. 

\bibitem{kickSPe}
J. D. Schnittman, A. Buonanno
Astrophys. J. Lett. 662 L63 (2007), 
J. D. Schnittman, A. Bounanno, J. R. van Meter, J. G. Baker, 
W. D. Boggs, J. Centrella, B. J. Kelly and S. T. McWilliams 
submitted to Phys. Rev. D, arXiv:0707.0301

\bibitem{kickBH}
M. J. Fitchett and S. D. Detweiler
Mon. Not. R. Astron. Soc. 211, 933 (1984), 
M. Favata, S. A. Hughes and D. E. Holz
Astrophys. J. Lett. 607, L5 (2004). 

\bibitem{PN} 
L. Blanchet, Living Rev. Relativity 9, (2006), 4.
http://www.livingreviews.org/lrr-2006-4, 
and references therein 

\bibitem{BH}
M. Sasaki and H. Tagoshi,
Living Rev. Relativity 6,  (2003),  6. 
http://www.livingreviews.org/lrr-2003-6, 
Y. Mino, M. Sasaki, M. Shibata, H. Tagoshi, T. Tanaka
Prog.Theor.Phys.Suppl. 128 (1997) 1-121, 
and references therein

\bibitem{QNM}
S. Chandrasekhar and S. L. Detweiler, 
Proc. R. Soc. London A344, 441(1975), 
S. L. Detweiler, 
Proc. R. Soc. London A352, 381(1977), 
S. L. Detweiler, 
Astrophys. J. 239, 292 (1980), 
E. W. Leaver,
Proc. R. Soc. London A402, 285(1986). 

\bibitem{QNMfit}
F. Echeverria, 
Phys. Rev. D40, 3194 (1989). 

\bibitem{QNMWKB}
V. Ferrari and B. Mashhoon, 
Phys. Rev. Lett. 52, 1361 (1984) 

\bibitem{QNMnu}
H. P. Nollert, 
Phys. Rev. D47, 5253 (1993)

\bibitem{akick}
M. Koppitz, D. Pollney, C. Reisswig, L. Rezzolla, 
J. Thornburg, P. Diener, E. Schnetter
Phys. Rev. Lett. 99, 041102 (2007), 
D. Pollney, C. Reisswig, L. Rezzolla, B. Szilagyi, 
M. Ansorg, B. Deris, P. Diener, E. N. Dorband, M. Koppitz, 
A. Nagar, E. Schnetter
Phys. Rev. D 76, 124002 (2007) 

\bibitem{plunge1}
M. Davies, R. Ruffini, W. H. Press and R. H. Press, 
Phys. Rev. Lett. 27, 1466 (1971), 
M. Davies, R. Ruffini and J. Tiomno, 
Phys. Rev. D5, 2932 (1972). 

\bibitem{plunge3}
T. Nakamura and M. Sasaki, 
Phys. Lett. A 89, 185 (1982), 
T. Nakamura and M. Haugan, 
Astrophys. J. 269, 292 (1982), 
T. Nakamura and M. Sasaki, 
Prog. Thore. Phys. 67, 1778 (1982). 

\bibitem{plunge5}
Y. Kojima and T. Nakamura
Phys. Lett. A 96, 335 (1983), 
Y. Kojima and T. Nakamura
Prog. Theore. Phys. 71, 79 (1984). 

\bibitem{plunge2}
M. Sasaki 
Prog. Theor. Phys. 69, 815 (1983).

\bibitem{plunge4}
K. Oohara and T. Nakamura
Prog. Theore. Phys. 70, 757 (1983). 

\bibitem{s_teu} E. D. Fackerell and R. G. Crossman, 
J. Math. Phys. 18, 1849 (1977). 

\bibitem{r_teu0} E. W. Leaver, 
J. Math. Phys. 27, 1238 (1986).

\bibitem{r_teu1} S. Mano, M. Suzuki and E. Takasugi, 
Prog. Theor. Phys. 95, 1079 (1996).

\end{thebibliography}



\end{document}